\documentclass[aps,prl,reprint,groupedaddress,nofootinbib,longbibliography]{revtex4-1}

\usepackage{amssymb,amsmath,bm,natbib}
\usepackage[dvipsnames]{xcolor}
\usepackage[colorlinks = true,
            linkcolor = Maroon,
            urlcolor  = Maroon,
            citecolor = Maroon]{hyperref}
\usepackage{microtype}
\usepackage{graphicx}
\usepackage{xspace}
\usepackage{bbm}
\usepackage{tikz}
\usetikzlibrary{decorations.markings}
\usetikzlibrary{decorations.pathmorphing}
\tikzset{snake it/.style={decorate, decoration=snake}}
\usetikzlibrary{matrix,calc,arrows.meta}
\usepackage{float}
\usepackage{physics}

\renewcommand{\textit}[1]{#1}

\usepackage{ifpdf}
\usepackage{tikz}
\usepackage[compat=1.1.0]{tikz-feynman}

\usepackage{xfrac}
\usepackage{fancyhdr}
\usepackage{mathbbol}
\usepackage{url}
\usepackage{color}
\usepackage{bbm}
\usepackage{bm}
\usepackage{slashed}
\usepackage{wasysym}
\usepackage{ytableau}
\usepackage{cancel}

\def\bl{\begin{equation}\begin{aligned}}
\def\el{\end{aligned}\end{equation}}
\def\beal{\begin{align}}
\def\eal{\end{align}}
\def\be{\begin{equation}}
\def\ee{\end{equation}}
\def\bpm{\begin{pmatrix}}
\def\epm{\end{pmatrix}}
\def\bsm{\begin{bmatrix}}
\def\esm{\end{bmatrix}}
\def\bea{\begin{eqnarray}}
\def\eea{\end{eqnarray}}

\def\tilde{\widetilde}

\newcommand{\OMIT}[1]{}

\begin{document}

\title{A New Look at  the $X$ Compositeness from its Lineshape}

\author{A. Carducci, G. Cianti, P. D'Annibali, D. Germani and A.D.~Polosa}
\affiliation{Dipartimento di Fisica, Sapienza Universit\`a di Roma, Piazzale Aldo Moro 2, I-00185 Rome, Italy}
\affiliation{INFN Sezione di Roma, Piazzale Aldo Moro 2, I-00185 Rome, Italy}

\date{\today}

\begin{abstract}
\noindent
{Available data on the $X(3872)$ lineshape have been fitted with a Flatt\`e parametrization derived from an effective Lagrangian in which $X$, $D$, and $D^*$ are all elementary fields.} A molecular $X$, i.e., a deuteron-like bound state of $D$ and $\bar{D}^*$, must instead be described by an effective theory where the $X$ field is initially absent or introduced as an auxiliary, unphysical,  field, yielding a lineshape parametrization that differs significantly from the Flatt\`e one. We demonstrate, in the analogous theory of nucleons with an auxiliary deuteron field, that this parametrization gives a fully consistent description of $np\to\text{deuteron}\to np$ scattering data, and motivate applying the same method to the $X$ lineshape. {The analysis proposed here is novel and has not previously been carried out on the available data.} 
 
\end{abstract}

\maketitle

\section{Introduction}

Whether the $X(3872)$ is interpreted as an elementary compact tetraquark or a molecular bound state of $D\bar D^*$ hinges on the choice of effective Lagrangian. If the Lagrangian contains $D$, $\bar D^*$ and $X$ as elementary fields, the $X$ is fundamental at the EFT scale. If instead the $X$ is dynamically generated, its pole arises from an infinite resummation of $D\bar D^*$ bubbles in the $s$-channel, and no fundamental $X$ field is present.

In~\cite{Kaplan:1996nv} it is shown that one can introduce an auxiliary quantum field $\Phi$ for the deuteron in a Lagrangian of nucleon fields $n,p$, but a consistent description of $np\to\Phi\to np$ requires $\Phi$ to have a wrong-sign kinetic term.\footnote{The field $\Phi$ is not physical. It is a bookkeeping device whose use is {\it restricted}  at tree level only to mediate $NN$ interactions. $\Phi$ cannot appear on external legs. Therefore loops with internal $\Phi$ particles are not included either because they would involve cuts with $\Phi$ external particles.   } Written in terms of $N=(p~n)$, with mass $M$, and dibaryon $\Phi$ of mass $2M$:
\begin{align}
\label{dbk}
\mathcal{L}_{\text{eff}}&=N^\dagger\!\left(i\partial_t+\frac{\nabla^2}{2M}\right)\!N
+\sigma\Phi^\dagger\!\left(i\partial_t+\frac{\nabla^2}{4M}-\Delta\right)\!\Phi\notag\\
&\quad-\left(y\Phi^\dag N N+\text{h.c.}\right)-\tfrac{1}{2}C\!\left(N^\dagger N\right)^2+\dots
\end{align}
with $\sigma=-1$  and $\Delta$ defining the bare mass of the deuteron $2M+\Delta$.\footnote{Shift the mass  $2M$ by $\Delta$ in the Klein-Gordon operator, and neglect $\Delta^2$: \mbox{$(\square -(2M)^2)\Phi-(4M\Delta)\Phi$}. In the non-relativistic limit \mbox{$\Phi\to e^{-i2Mt}\Psi$} and $(\square-(2M)^2) \Phi \to e^{-i(2M)t}(\nabla^2+2i(2M)\partial_t)\Psi$. Renormalize $\Psi$  by  $\sqrt{4M}$ to obtain~\eqref{dbk}.} Replacing $n,p\to D,\bar D^*$ and $\Phi\to X$ generates a parametrization for the $X$ lineshape distinct from Flatt\`e's. The latter, used by LHCb, implicitly treats $X$ as fundamental on equal footing with $D,\bar D^*$ — a good Flatt\`e fit therefore tests self-consistency of the elementary hypothesis, not the molecular picture. {Alternative methods to test the molecular nature of the $X(3872)$ have been developed in \cite{Braaten:2015tga,Braaten:2020nmc}}. 

In~\cite{landau,Weinberg:1965zz} the coupling $g_B$ of a loosely bound state to its constituents is given in terms of the reduced mass $m$ and binding energy $B$ as
\be
g_B^2=\frac{2\pi}{m}\sqrt{\frac{2B}{m}}.
\label{landau}
\ee
We first show that, with the auxiliary field $\Phi$ and the relation between $y$ and $g_B$, we can reproduce, with good approximation, the measured deuteron effective range
\be
r_0^{\rm exp}=+1.73~{\rm fm},
\label{expr0}
\ee
the canonical signature of a bound state~\cite{Weinberg:1965zz}. We then turn to $X(3872)$.  
This suggests that $g_B$ can also be used as a benchmark for the $X$ study.

\section{The Deuteron Effective Range}
In the non-relativistic limit the effective coupling of the deuteron to nucleons is
\be
\mathcal{L}_I=-y\Phi_i^\dag\,\epsilon_{ab}\,N_a^T\sigma_2\sigma_i N_b.
\label{unoint0}
\ee
{here $N_a$ is the two-component nucleon field carrying an isospin index $a=1,2$ ($N_1=p$, $N_2=n$) and an implicit spin index. The deuteron field $\Phi_i$ is a vector under spatial rotations where $i=1,2,3$ labels its components.}
The iso-singlet projection rescales $\epsilon_{ab}\to\epsilon_{ab}/\sqrt{2}$, and the spin-triplet components are read off via
\be
\frac{-i}{\sqrt{2}}(\bm e^*_{(\lambda)}\cdot\sigma_2\bm\sigma),
\ee
in the spherical basis of polarization vectors.
Defining $\mathcal{L}_I$ such that the one-$\Phi$-exchange diagram yields the same  Feynman rule as in~\cite{Kaplan:1996nv}
\begin{align}
    \begin{tikzpicture}[baseline={([yshift=-0.6ex]current bounding box.center)}, scale=0.7, transform shape]
        \begin{feynman}
        \vertex (i1) at (0,1) {\(n\)};
        \vertex (i2) at (0,-1) {\(p\)};
        \vertex (v1) at (1,0);
        \vertex (v2) at (2.5,0);
        \vertex (f1) at (3.5,1) {\(n\)};
        \vertex (f2) at (3.5,-1) {\(p\)};
        \vertex at (1.8,0.3) {\(\Phi\)};
        \diagram*{
            (i1) -- (v1),
            (i2) -- (v1),
            (v1) -- [very thick] (v2),
            (v2) -- (f1),
            (v2) -- (f2),
        };
        \end{feynman}
    \end{tikzpicture}\quad\to\quad y^2\frac{\sigma}{E-\Delta},
    \label{eq:diagrams}
\end{align}
allows to use all the results obtained therein. 
Decomposing the scattering state as 
\be
\lvert np\rangle=\tfrac{1}{\sqrt{2}}\!\left(\left\lvert\tfrac{np-pn}{\sqrt{2}}\right\rangle+\left\lvert\tfrac{np+pn}{\sqrt{2}}\right\rangle\right),
\ee
in order to project onto the iso-singlet, $S=1$ component, we have 
\bea
&&\!\!\left\langle d\!\left|i\frac{y}{\sqrt{8}}\Phi_i^\dag N^T\tau_2\sigma_2\sigma_i N\right|\!\left(\frac{N^Ti\tau_2 N}{2}\right)_{\!\!S=1}\right\rangle=\notag\\
&&=i\frac{y}{\sqrt{2}}e_i^*\chi_n^T\sigma_2\sigma_i\chi_p,
\label{landauc}
\eea
{where $\chi_n$ and $\chi_p$ are the two-component Pauli spinors of the external proton and neutron and $(i\tau_2)_{ab}=\epsilon_{ab}$ acts on isospin indices.}
Using $\sum_{\rm pol}e_i^*e_j=\delta_{ij}$, the spin trace~\footnote{where in the spin-triplet configuration either all the spinors in the trace are up, or all are down, or $\uparrow\uparrow\downarrow\downarrow+\downarrow\downarrow\uparrow\uparrow$  (the mixed configurations $\uparrow\downarrow\downarrow\uparrow, \downarrow\uparrow\uparrow\downarrow$ giving trace equal to zero).}
\be
\sum_i\!\langle\sigma_2\sigma_i(\chi_p\chi_p^T)\sigma_i\sigma_2(\chi_n\chi_n^T)\rangle=2
\label{trace}
\ee
allows to recover the diagrammatic Eq.~\eqref{eq:diagrams}.

The matrix element $|\langle d|V|(NN)_{S=1,I=0}\rangle|^2$ has to  be proportional to $g_B^2$~\cite{Esposito:2025hlp}. Here $d$ is the deuteron bound state destroyed by $\Phi$, and $NN$ is the constituents' state as in~\eqref{landauc}.   Hence from~\eqref{landauc} we  conclude
\be
y=\lambda\sqrt{2}\,g_B
\label{main}
\ee
with the proportionality factor $\lambda$ determined explicitly in~\eqref{LambdaAlpha}. 

The Lippmann-Schwinger equation with kernel~\footnote{The optical theorem (unitarity) is enforced by $NN$ bubbles only, no $\Phi$ in the loops.}
\be
V_{\rm eff}(p)=C+\frac{\sigma y^2}{E-\Delta},\qquad E=\frac{p^2}{2m},
\ee
corresponding to the sum of the two Feynman diagrams above, yields
\be
\mathcal{A}_{\text{eff}}=-\frac{C(E-\Delta)+\sigma y^2}{E-\Delta+i\frac{Mp}{4\pi}\!\left(C(E-\Delta)+\sigma y^2\right)},
\label{KF}
\ee
where for $E>0$
\be
\sqrt{-2mE-i\epsilon}=-i|\bm p|=-ip=-i\sqrt{2mE}.
\label{detsqrt}
\ee

Comparing with the non-relativistic-QM result for $\mathcal{A}_{\text{eff}}$ in the regularized $\delta$-potential~\footnote{In EFT, short-distance physics at $r<1/\Lambda$ is encoded in contact operators with unknown coefficients; the $\delta$-function is the lowest-order one. Following~\cite{Kaplan:1996nv} we adopt the zero-range approximation, dropping the Yukawa: the $\delta$-shell models the outcome of binding, not its mechanism.}
\begin{equation}
V(r)=-g\frac{\Lambda}{M}\delta\!\left(r-\frac{1}{\Lambda}\right),
\label{deltapot}
\end{equation}
where $g\equiv\Lambda M v/(4\pi)$ and $V(r)$ derives from regularizing at $R=1/\Lambda$ the potential $V(r)=-v\delta^3(r)=-v\tfrac{\delta(r)}{4\pi r^2}$. Matching the discontinuity of the logarithmic derivative at $r=R$ we find,~\footnote{Explicitly $1/v=(M/4\pi R)(1-\exp(-\sqrt{4MB}/\Lambda))/(\sqrt{4MB}/\Lambda)$.}
\be
g=\frac{\sqrt{4MB}/\Lambda}{1-\exp(-\sqrt{4MB}/\Lambda)},
\label{fg}
\ee
so $g\geq 1$ in the presence of a bound state. With $\sigma=-1$ the matching conditions are (the EFT parameters on the lhs and the QM parameters on the rhs)~\cite{Kaplan:1996nv}:
\begin{align}
y^2=\frac{(1+g)}{g}\frac{3\pi\Lambda}{4m^2},\;\;
C=\frac{\pi}{m\Lambda},\;\;
\Delta=-\frac{(1-g)}{g}\frac{3\Lambda^2}{4m},
\label{tre}
\end{align}
where $m$ is the $pn$ reduced mass, and
\begin{equation}
\begin{cases}
a^{-1}=\dfrac{2\pi}{m}\dfrac{\Delta}{C\Delta+y^2}=-\!\left(\dfrac{1-g}{g}\right)\!\Lambda,\\[6pt]
r_0=\dfrac{2\pi}{m^2}\dfrac{y^2}{(C\Delta+y^2)^2}=\!\left(\dfrac{1+g}{g}\right)\dfrac{2}{3\Lambda}.
\end{cases}
\label{20}
\end{equation}
Inserting the Ansatz~\eqref{main}, with $\lambda=1$,  into~\eqref{tre} and using~\eqref{20} with $a^{\rm exp}=5.4194\pm0.0020$~fm, $m=469.5$~MeV, $B=2.22$~MeV gives~\footnote{With $\Lambda$ from~\eqref{glan}, Eq.~\eqref{fg} returns $g\simeq 1.36$, consistent with~\eqref{glan}.}
\begin{equation}
g=1.35,\qquad\Lambda=139\,\text{MeV},
\label{glan}
\end{equation}
consistent with $g>1$ (and $\Delta\ll M$) and a prediction for the effective range $
r_0^{\rm th}=+1.64\,\text{fm}$. The positive $r_0$ sign, necessary for a bound state~\cite{smorod,lqm3-2,Bethe,Esposito:2021vhu},  is a direct consequence of $\sigma=-1$ whereas $\Lambda\simeq m_\pi$ identifies the first integrated-out degree of freedom.

We can be more accurate than this, and with the further (almost inconsequential, see the Appendix) simplification $C=0$, compute the pole in  Eq.~\eqref{KF} at $y=y^\star$ where  
\be y^\star = \lambda \sqrt{2}g_B,\quad \lambda=\sqrt{\frac{\Delta+B}{4B}}
\label{landa}
\ee
and compute the residue at the pole finding the shallow bound state expression\footnote{{ Following~\cite{Kaplan:1996nv}, $f(E)$ is obtained by multiplying $A_{\rm eff}$ by $M/4\pi=m/2\pi$ which is the standard factor in front of the shallow bound state scattering amplitude, see Eq. (3.21) in~\cite{Esposito:2025hlp}. } }
{
\be
A_{\rm eff}(E)=-\frac{\left(\alpha\sqrt{2} g_B\right)^2}{E+B}
\ee
where the correction factor to the Landau coupling\footnote{confront with the $(1-Z)$ in Eq. (3.24) of~\cite{Esposito:2025hlp}} is determined by 
\be
\alpha=\frac{\lambda}{\sqrt{2\lambda^2-1}}
\ee
consistent with $\lambda=1\Rightarrow \alpha=1$.
}
{The expressions for $\lambda$ and $\alpha$ together with $\Delta=my^2/(2\pi a)$ (see~\eqref{20}) and the numerical values of $a^{\rm exp}$ and $B$, give two  approximately equal to one figures 
\be
\lambda=1.12, \quad
\alpha
= 0.91
\label{LambdaAlpha}
\ee}
{\it showing that, as expected,  the shallow bound state formula for the deuteron  is essentially $1/(E+B)$ times the square of the $g_B$ coupling, corrected by the Clebsch-Gordan factor}.

{We can use these values in~\eqref{20} to obtain $r_0$ as in:
\begin{equation}
 r_0=\frac{2\pi}{m^2y^2}=+1.73\text{ fm}
 \label{R0th},
\end{equation}
in perfect agreement with the experimental value.}
This shows how the Landau coupling is wired in the amplitude formula with $\sigma=-1$.  {In the molecular hypothesis, the $X(3872)$ is a bound state even shallower than the deuteron. The coupling of the $X$ to its components in the corresponding amplitude $f_+(E)$ (see next Section) takes naturally the Landau coupling value $g_B$. This places  a inevitable constraint on the experimental analysis.}

In the last Section we will see how this construction  enables a consistent description of the deuteron and motivates the pursuit of the same method to study the $X$: use $g_B$ as a starting and reference value for the $y$ coupling in the  model fitting the $X$ lineshape.

\paragraph{The $\delta$-shell model for the $X$.}
{Unlike the deuteron, where pion-mediated tensor forces drive the binding, pion exchange alone in $D\bar D^*\to D\bar D^*$ cannot form a bound state~\cite{Baru:2015nea}.\footnote{In the deuteron, the pion-induced $\delta^3(\bm x)$ term is repulsive, factorized by $\langle\bm\sigma_1\cdot \bm\sigma_2\rangle=2S(S+1)-3=1$ and $\langle\bm\tau_1 \cdot \bm\tau_2\rangle=2I(I+1)-3=-3$, yielding altogether a $+\delta^3(\bm x)$ term, accompanied by a real Yukawa potential $\propto-3\mu^2 e^{-\mu r}/r$ rather than a complex potential as in~\eqref{24}.} This conclusion was also reached in Ref.~\cite{Esposito:2023mxw} within a zero-range approximation, employing a single delta interaction and accounting for the finite $D^*$ width.} The Fourier-transformed pion propagator gives
\bea
(V_w)_{ij}&=&-\frac{\beta}{2}\!\int\!\frac{q_iq_j\,e^{i\bm q\cdot\bm r}}{\bm q^2-\mu^2-i\epsilon}\frac{d^3q}{(2\pi)^3}\notag\\
&=&-\frac{\beta}{6}\!\left(\delta^3(r)+\mu^2\frac{e^{i\mu r}}{4\pi r}\right)\!\delta_{ij},
\label{24}
\eea
with
\be
-\mu^2=-q_0^2+m_\pi^2=-(m_{D^*}-m_D)^2+m_{\pi^0}^2\simeq-(44~\text{MeV})^2,\notag
\ee
and $\beta=24\pi\alpha/\mu^2$, $\alpha=5\times10^{-4}$. The full $V_w=(V_w)_{ij}e_i^\lambda(\bm p)\bar e_j^{\lambda'}(\bm p')$ cannot bind for any polarizations, and the $\delta$-shell coupling is
\be
g=\frac{\alpha}{\mu^2}\,2m\Lambda,\qquad m=967\,\text{MeV}.
\label{gmol}
\ee
Achieving $g\geq 1$ requires $\Lambda\gtrsim 2$~GeV, i.e.\ $R\lesssim 0.1$~fm — distances at which mesons are not pointlike.

This requires to model the molecular binding in the $X$ by an extra short-range $\delta$-potential~\cite{Baru:2015nea,Esposito:2023mxw} and introduce an auxiliary field for $X$, with $\sigma=-1$, to parallel  the deuteron treatment. In the next section we upgrade~\eqref{KF} to include the second ($D^+\bar D^{*-}$) threshold.

\section{The $f_+$ parametrization for the molecular $X$}

The universal $S$-wave amplitude at low energies is\footnote{$k$ is the recoiling momentum of the scattering pair.}
\begin{align}
f=\frac{1}{-1/a+\tfrac{1}{2}r_0 k^2-ik}.
\label{amplitude}
\end{align}
A negative $r_0$ indicates a non-zero field-renormalization $Z$ for $X$ (elementary), a positive one indicates  a bound state~\cite{Weinberg:1965zz}. With insufficient data near $k\simeq 0$, the analysis must be extended to larger $k$, where the charged $D^+\bar D^{*-}$ threshold at $\delta\simeq 8.2$~MeV becomes relevant.

The non-relativistic Flatt\`e amplitude~\cite{Flatte,Artoisenet:2010va,Hanhart:2007yq} (compact $X$) is\footnote{Up to the $(2\pi)^4$ factor. The $\Im(a)<0$ with $\Gamma_r>0$.}
\begin{equation}
f_-\equiv\frac{-\tfrac{mg^2}{2\pi}}{E-\Delta+i\tfrac{mg^2}{2\pi}\!\left[\sqrt{2mE}+i\sqrt{-2m(E-\delta)}\right]+\tfrac{i\Gamma_r}{2}},
\label{F}
\end{equation}
with $m\simeq 967$~MeV the $D\bar D^*$ reduced mass, $\Delta$ the Flatt\`e mass parameter, and $\Gamma_r$ a remaining inelastic width.\footnote{The $mg^2/\pi \, \sqrt{2mE}$ term is the non-relativistic decay width of a spin-1 particle of mass $\sqrt{s}=m_D+m_{D^*}+E\simeq m_X+E$, and $s\simeq m_X^2+2m_X E$ for $E\ll m_D, m_{D^*}$. The same result follows from the optical theorem $\Im f(0)=(k/2\pi)\sigma_T$ applied to the Flatt\`e propagator.} 

The molecular amplitude is instead obtained from~\eqref{KF} by setting $\sigma=-1$ and absorbing $C$ (see Appendix):~\footnote{For a treatment with also the iso-triplet see~\cite{Ji:2025hjw}.}
\bea
f_+(E)&=&\frac{M}{4\pi}\,\mathcal{A}_{\rm eff}=\\
&=&-\frac{M}{4\pi}\,\frac{-y^2}{E-\Delta-i\tfrac{Mp}{4\pi}y^2}=\frac{\tfrac{my^2}{2\pi}}{E-\Delta-i\tfrac{my^2}{2\pi}\sqrt{2mE}},\notag
\eea
which, augmented with the charged-threshold imaginary part for $E<\delta$ and the inelastic width, becomes
\begin{equation}
f_+\equiv\frac{\tfrac{my^2}{2\pi}}{E-\Delta-i\tfrac{my^2}{2\pi}\!\left[\sqrt{2mE}+i\sqrt{-2m(E-\delta)}\right]-\tfrac{i\Gamma_r}{2}}.
\label{K}
\end{equation}
In $f_-$ the coupling $g$ is purely phenomenological. A consistent fit of LHCb data with $f_+$ should also return $y$ compatible with the Landau coupling value~\eqref{main}.


Comparing~\eqref{F} and~\eqref{K} with~\eqref{amplitude} yields $1/a^{\pm}=\pm 2\pi\Delta / (m\,g_{\pm}^2)-\sqrt{2m\delta} + i\pi /(m g_{\pm}^2) \Gamma_r$ and 
\begin{equation}
r_0^{\pm}=\pm\frac{2\pi}{m^2g_{\pm}^2}-\frac{1}{\sqrt{2m\delta}}.
\end{equation}
Without the charged threshold, $r_0^+>0$ (while $r_0^-<0$), with
\be
r_0^+=\frac{1}{\sqrt{2mB}}-\frac{1}{\sqrt{2m\delta}},
\ee
so $r_0^+>r_0^{\rm min}=-\! 1/\sqrt{2m\delta}=-1.58$~fm. Here $g_+=y=g_B$ and $g_-$ is the phenomenological coupling of a compact $X$ in~\eqref{F}. For the deuteron (no second threshold, $y=\sqrt{2}g_B$), $r_0^+=1/\!\sqrt{8mB}$: with $m=938/2$~MeV and $B=2.2$~MeV one finds $r_0^+=2.1$~fm, a result obtained without the factor $\lambda$ but still consistent with~\eqref{R0th} and~\eqref{expr0}.
On the basis of available LHCb data~\cite{LHCb:2020xds}, the effective range of the $X$ is $r_0^-=-5.34$~fm (large and negative)~\cite{Esposito:2021vhu}, but some proposal have been formulated to revise this result bringing it towards zero, e.g.~\cite{r0_isospin}.  


The branching ratio $\mathcal{B}_X\equiv\mathcal{B}(X\to DD\pi)$ is measured more accurately than the total width $\Gamma_X$, and $\mathcal{B}_X\Gamma_X=\Gamma(X\to DD\pi)$ scales with $g_B^2$~\eqref{landau}. Fixing $B$ and $\mathcal{B}_X$ thus sets a conservative lower bound on $\Gamma_X$ for a molecular $X$, as shown in~\cite{Polosa:2015tra}. We take this bound as one benchmark to simulate $|f_+|^2$ pseudo-data and refit them with both $f_+$ and $f_-$.

\begin{figure}[htbp]
    \centering
    \includegraphics[width=0.45\textwidth]{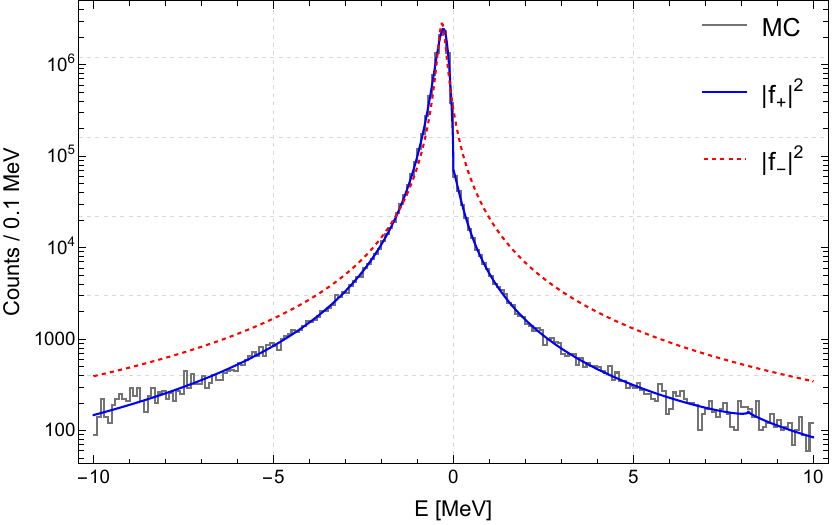}
   \caption{$D\bar D^*\to X(3872) \to D\bar D^*$ simulated events with
    $B = 200$~keV and $\Gamma_X = 480$~keV.
    Parameters of $f_+$: $\Delta = 2.85$~MeV, $\Gamma_r = 20$~keV,
    $y= 0.012~\text{MeV}^{-1/2}$, fixed by the pole condition at $E = -B - i\Gamma_X/2$.
    Grey histogram: Monte Carlo sample from $|f_+|^2$ ($10^6$ events, 100~keV bins, shown for visualization only).
    red curve: best fit with $|f_+|^2$; red curve: best fit with $|f_-|^2$
    (unbinned maximum-likelihood). In the very low energy region $f_{\pm}$ collapse on the same universal $f$ in~\eqref{amplitude}.}
    \label{fig:yourlabel}
\end{figure}
LHCb {\it data have so far been fitted exclusively with $f_-$} \cite{LHCb:2020xds},  and the quality of that fit raises the question of whether $f_+$ --- with $y$  constrained by the Landau coupling as in~\eqref{main} --- can perform equally 
well. This constitutes the discriminating test of the molecular hypothesis  (see Fig.~\ref{fig:yourlabel}): if the $X$ is molecular, a good fit with $f_+$,  with the coupling of the $X$ to its constituents fixed at the natural Landau value, is  unavoidable.

\section{Deuteron parameters  in $np$ scattering from $f_+(E)$}
\begin{figure}[h]
    \centering
    \includegraphics[width=0.45\textwidth]{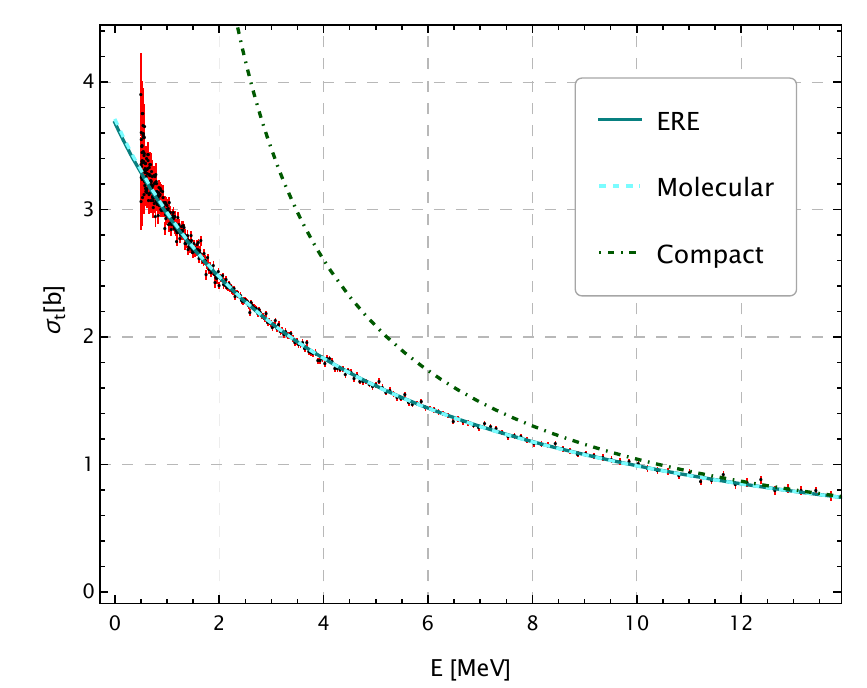}
   \caption{Fit on the $np$ scattering data in triplet channel \cite{Clement:1972rut}. The fit has been performed with three different models: the compact one $(4\pi|f_-^{np}|^2)$, the molecular one $(4\pi|f_+^{np}|^2)$ and the Effective Range Expansion (ERE) \eqref{amplitude}\,$ (4\pi|f|^2)$. Differently from Fig.~\ref{fig:yourlabel} tails are better fit than the peak region, due to the relative weight of statistical errors. }
    \label{fig:Fit_Deuteron}
\end{figure}
With the data obtained by \texttt{EXFOR} \cite{Otuka:2014wzu} from the experimental work done in \cite{Clement:1972rut} we can show visibly that the molecular model correctly describes the deuteron state, as displayed in Figure \ref{fig:Fit_Deuteron}. The fitted cross section is obtained from the unpolarized $np$ scattering process through the formula
\begin{equation}
    \sigma_{\text{tot}}(k)=\frac{3}{4}\sigma_{t}(k)+\frac{1}{4}\sigma_s(k)\,.
\end{equation}
Here $k=\sqrt{2m_{np}E}$ is the relative momentum of the $np$ pair, while $\sigma_t$ ($\sigma_s$) refers to the spin-triplet (-singlet) channel. In order to extract the data associated with the triplet channel (the one forming the deuteron bound state), we modeled the spin-singlet cross section using the effective range expansion in Eq. \eqref{amplitude}
\begin{equation}
    \sigma_s(k)=4\pi\left|\frac{1}{-\frac{1}{a_s}+\frac{1}{2}{r_0}_sk^2-ik}\right|^2,
\end{equation}
with $a_s=-23.711\,\text{fm}$ and ${r_0}_s=2.76\,\text{fm}$ taken from \cite{Koester:1990ev}. For the spin-triplet channel, we used the expansion \eqref{amplitude} and the amplitudes $f_-$ (compact) and $f_+$ (molecular), modified accordingly for the deuteron case which is stable and has no second threshold (i.e. Eq. \eqref{KF} setting $C=0$)
\begin{equation}
f_\pm^{np}=\frac{\pm\tfrac{m_{np}}{2\pi}y_\pm^2}{E-\Delta_\pm\mp i\tfrac{m_{np}}{2\pi}y_\pm^2\sqrt{2m_{np}E}}\,.
\label{eq:f_fit_pm}
\end{equation}
The results of these three non-linear fit are the following:
\begin{align}
    &a_t=5.435(6)\,\mathrm{fm},\qquad {r_0}_t=1.71(1)\,\mathrm{fm};\\
    &y_+=0.0573(2)\,\mathrm{MeV}^{-\frac{1}{2}},\qquad \Delta_+=8.90(7)\,\mathrm{MeV}\label{eq:yp_deltap}\\
    &y_-\simeq150\,\mathrm{MeV}^{-\frac{1}{2}},\qquad \Delta_-\simeq-62\,\mathrm{MeV}.
\end{align}
As one can easily see, the effective range expansion and the molecular model can fit the experimental data successfully, in contrast with the compact model for which the fit does not converge to a stable minimum (hence we avoided reporting the error on the bestfit parameters in this case). Furthermore, one can check that the best fit values obtain are in good agreement with the existing experimental data for $a_t$ and ${r_0}_t$ \cite{Koester:1990ev}, while for the molecular model the best fit value must be compared with $y^\star\simeq1.12\sqrt{2}g_B=0.0571\textrm{ MeV}^{-\frac{1}{2}}$ of Eq.~\eqref{landa}, which is perfectly in accordance with the fitting parameter.\\
The phase of $f_+^{np}$ decreases with increasing energy, corresponding to a clockwise phase rotation in the Argand plane, in agreement with the low-energy ${}^3S_1$ phase shifts extracted from $np$ scattering data \cite{Stoks1993,NavarroPerez2013}.

\paragraph{Analytical structure of $f_\pm$.}
{Using $y_+$ and $\Delta_+$ from Eq.~\eqref{eq:yp_deltap}, the denominator of $f^{np}_+$ exhibits two poles. The first pole at $E_1 \simeq -2\,\text{MeV}$ ($|k_1| \simeq 43\,\text{MeV}$) corresponds to the physical near-threshold state ($|k_1| \sim 1/|a|$, the deuteron).} The second pole at $E_2 \simeq -40~\text{MeV}$ corresponds instead to 
$|k_2| \simeq 199~\text{MeV} \sim 2/r_0$, lying beyond the region of 
validity of the effective theory, as delimited by the cutoff 
$|k| \ll \Lambda \sim 1/r_0$. This is a well-known artifact of effective 
range expansions (see, e.g., \cite{Bedaque:1999vb}), arising from the 
truncation of the denominator at order $k^2$.

{Both the molecular amplitude in Eq.~\eqref{K} and the Flatt\'e amplitude in  Eq.~\eqref{F} inherit this spurious pole structure. A non-zero $\Gamma_r$  shifts the unphysical pole into the upper ($f_+$) or lower ($f_-$) complex  energy half-plane, mimicking a resonance in the latter case. As long as the 
physical state lies close to threshold, however, the spurious poles remain well beyond the EFT breakdown scale and leave the near-threshold predictions unaffected.}

\section{Conclusions}


The Flatt\`e amplitude $f_-(E)$, used by LHCb, implicitly takes $X$ as an elementary field on equal footing with $D,\bar D^*$. The negative $r_0^-$ that follows is  negative by construction, and the Weinberg compositeness criterion as applied to LHCb results is circular: it reproduces the assumption rather than testing it.

To test the molecular hypothesis independently, data should be fitted with $f_+(E)$, derived from the auxiliary-field Lagrangian of~\cite{Kaplan:1996nv} with $\sigma=-1$ and $y$ set by the Landau coupling $g_B$ in~\eqref{landau}. We have validated the framework on the deuteron: it  reproduces $r_0^{\rm exp}=+1.73$~fm with no free parameters beyond $a$ and $B$.  We have checked that $f_+$, differently from $f_-$, fits well deuteron data. 

The pure molecular hypothesis can be supported if $f_+$ (i)~provides a good fit to LHCb data, (ii)~with coupling compatible with the Landau value; Eqs.~\eqref{main} and~\eqref{landa}.

Recent literature pursues the short-distance compact component of the $X$~\cite{Maiani:2004vq} via Born--Oppenheimer methods~\cite{BOold1,BOold2,BOold3,Germani:2025mos,BOrattz,brambi0,Brambilla:2024thx,Braaten:2024tbm}. A renewed experimental effort to test the purely molecular picture along the lines followed in this paper would help confront the ongoing paradigm shift on the $X(3872)$ and related exotic hadrons.

\begin{acknowledgments}
{\it  Acknowledgments:  The work of PD is supported by the MUR FIS2 Advanced Grant ET-NOW (CUP: B53C25001080001) and by the INFN TEONGRAV research project. We acknowledge financial support under the program ``Progetti di Rilevante Interesse Nazionale'' (PRIN 2022), published on 2.2.2022 by the Italian Ministry of University and Research (MUR), Project Title ``The X(3872) files'' -- CUP B53C24006500006 -- Grant Assignment Decree No. 20429 adopted on 6.11.2024 by the Italian Ministry of University and Research (MUR).}
\end{acknowledgments}

\appendix
\section{Appendix. Hubbard--Stratonovich Transformation: the case of the Deuteron}

We show how to introduce an auxiliary dibaryon field $\Phi$ to reorganize the contact two-nucleon interactions. Starting from the Lagrangian in~\cite{Kaplan:1996nv} and projecting it onto the deuteron quantum numbers \footnote{We adopt the notation $O_N^\dagger O_N=|O_N|^2$}:
\bea
\widetilde{\mathcal L}_{\rm eff}
&=&N^\dagger\!\left(i\partial_t+\frac{\nabla^2}{2M}\right)\!N
-\tfrac{1}{2}\widetilde C|N^T P_i N|^2 \notag\\
&-&\tfrac{1}{2}\widetilde C_2|N^T P_i \nabla N|^2+\dots,
\label{iniziale}
\eea
where $P_i=\frac{1}{\sqrt{8}} \sigma_2 \sigma_i \tau_2$ projects onto the deuteron quantum numbers. We insert into the path integral the Gaussian identity
\be
\int\mathcal{D}\mu\,
e^{\,i\!\int d^{4}x\,\widetilde{\mathcal{L}}_{\rm eff}}\,
\int \mathcal{D}\Phi\,\mathcal{D}\Phi^{\dagger}\,e^{\,i\!\int d^4x\,d^4y\,(\Phi^\dag-F^\dag)K_0(\Phi-F)},
\ee
where $F$ is an arbitrary functional of nucleon fields and  $\Phi$, the auxiliary field. The kinetic operator is 
\begin{equation}
K_0=\sigma\!\left(i\partial_t+\frac{\nabla^2}{4M}-\Delta\right),\qquad\sigma=\pm 1.
\end{equation}
Expanding the Gaussian term,
\begin{align}\notag
\mathcal L_{\rm eff}&=N^\dagger\!\left(i\partial_t+\tfrac{\nabla^2}{2M}\right)\!N
-\tfrac{1}{2}\widetilde C|N^T P_i N|^2 \notag-\\&-\tfrac{1} {2}\widetilde C_2|N^T P_i \nabla N|^2 +F^\dag K_0 F+\notag \\
&+\Phi^\dag K_0\Phi-\Phi^\dag K_0 F-F^\dag K_0\Phi+\dots
\end{align}
Choosing $F=y\,K_0^{-1}N^T P_i N$ generates the trilinear coupling in~\eqref{dbk}, and\footnote{$K_0^{-1}$ follows from inverting $K_0$ as a formal geometric series in $D/\Delta$ with $D\equiv i\partial_t+\nabla^2/(4M)$: each higher-derivative insertion is suppressed by one extra power of $\Delta$, reflecting the heaviness of the dibaryon relative to the low-energy scales probed by $D$.}
\be
F^\dag K_0 F=y^2(N^T P_i N)^\dag K_0^{-1}(N^T P_i N),
\ee
with
\be
K_0^{-1}=\sum_{n=0}^{\infty}\frac{\sigma}{\Delta^{n+1}}\!\left[i\partial_t+\frac{\nabla^2}{4M}\right]^n.
\ee
The contact interactions can be removed by imposing matching conditions at each order of $K_0^{-1}$. At $n = 0$:
\be
y^2(N^T P_i N)^\dag \frac{\sigma}{\Delta}(N^T P_i N)-\tfrac{1}{2}\widetilde C|N^T P_i N|^2=0 
\label{38}
\ee
Therefore:
\be
\widetilde C = \frac{2 \sigma y^2}{\Delta}
\ee
On the other hand, if we had $\tilde C=0$ in~\eqref{iniziale}, the HS transformation would give the four-fermion interaction with $-\tfrac{1}{2}C$ and $ C>0$ as in~\eqref{dbk},  if $\sigma=-1$. This matches an attractive $\delta$ potential.  In this sense, $\Phi$ is not a physical particle but a stand-in for the interaction it generates.
For the $n=1$ term:
\be
y^2(N^T P_i N)^\dag \frac{\sigma}{4 M \Delta^2}\nabla^2 (N^T P_i N)-\tfrac{1} {2}\widetilde C_2|N^T P_i \nabla N|^2 =0
\ee
Integrating the first term by parts:
\be
 -\int d^3x\, \frac{\sigma y^2}{4\,\Delta^2 M}\nabla(N^TP_i N)^\dagger \cdot \nabla(N^TP_i N)
  \label{41}
\ee
and
\begin{equation}
  \nabla(N^T P_i N) = (\nabla N)^T P_i\, N + N^T P_i\, (\nabla N)= 2\, N^T P_i\, (\nabla N),\notag
\end{equation}
since  $N$ are Grassmann fields  and $\sigma_2\sigma_i^*\sigma_2=-\sigma_i$.
Analogously for the Hermitian conjugate in~\eqref{41}, leading to $-\int d^3x \frac{\sigma y^2}{  \Delta^2 M}|N^T P_i \nabla N|^2$.
Therefore in addition to~\eqref{38}:
\begin{equation}
-\frac{y^2\sigma}{ M \Delta^2}|N^T P_i \nabla N|^2-\tfrac{1} {2}\widetilde C_2|N^T P_i \nabla N|^2=0,
\end{equation}
which is solved by
\begin{equation}
 \widetilde C_2 = -\frac{2 y^2\sigma}{ M \Delta^2}.
\end{equation}
Following~\cite{Kaplan:1996nv}, the leading $(NN)^2$ interaction is retained for matching to the $\delta$-potential, while the derivative term is reabsorbed, recovering~\eqref{dbk}.

\bibliography{ref.bib}

\end{document}